# Applying the Principles of Augmented Learning to Photonics Laboratory Work

U.H.P. Fischer[1] (member IEEE), Matthias Haupt[2], Christian Reinboth[3], Jens-Uwe Just[4]

[1]Harz University, Friedrichstraße 57-59, D-38855 Wernigerode
(49) 3943 659 105, (49) 3943 659 399 (fax), ufischerhirchert@hs-harz.de

Harz University, Friedrichstraße 57-59, D-38855 Wernigerode
(49) 3943 659 368, (49) 3943 659 399 (fax), mhaupt@hs-harz.de

[3]HarzOptics Photonics Research GmbH, Dornbergsweg 2, D-38855 Wernigerode
(49) 3943 935 615, (49) 69 1539 6333 858 (fax), creinboth@harzoptics.de

[4]HarzOptics Photonics Research GmbH, Dornbergsweg 2, D-38855 Wernigerode
(49) 3943 935 615, (49) 69 1539 6333 858 (fax), jjust@harzoptics.de

**Abstract**: Most modern communication systems are based on opto-electrical methods, wavelength division multiplex (WDM) being the most widespread. Likewise, the use of polymeric fibres (POF) as an optical transmission medium is expanding rapidly. Therefore, enabling students to understand how WDM and/or POF systems are designed and maintained is an important task of universities and vocational schools that offer education in photonics.

In the current academic setting, theory is mostly being taught in the classroom, while students gain practical knowledge by performing lab experiments utilizing specialized teaching systems. In an ideal setting, students should perform such experiments with a high degree of autonomy. By applying the principles of augmented learning to photonics training, contemporary lab work can be brought closer to these ideal conditions.

This paper introduces „OPTOTEACH", a new teaching system for photonics lab work, designed by Harz University and successfully released on the German market by HarzOptics. OPTOTEACH is the first POF-WDM teaching system, specifically designed to cover a multitude of lab experiments in the field of optical communication technology.

It is illustrated, how this lab system is supplemented by a newly developed optical teaching software - „OPTOSOFT" - and how the combination of system and software creates a unique augmented learning environment. The paper details, how the didactic concept for the software was conceptualised and introduces the latest beta version. OPTOSOFT is specifically designed not only as an attachment to OPTOTEACH, it also allows students to rehearse various aspects of theoretical optics and experience a fully interactive and feature-rich self-learning environment.

The paper further details the first experiences educators at Harz University have made working with the lab system as well as the teaching software. So far, the augmented learning concept was received mostly positive, although there is some potential for further optimisation concerning integration and pacing of various interactive modules.

## 1 Introduction

The demands on digital high-speed data communication equipment are increasing permanently and so are the demands on the maximum bandwidth of transmission media [Na00]. Modern communication systems need high-speed optical transmitters and receivers for Terabit data transmission rates. Most of these communication systems are based on advanced opto-electrical methods like wavelength division multiplex (WDM), which is one of the most widely used methods. Likewise, the use of polymer optical fibres (POF) as an optical transmission medium is expanding rapidly.

The POF is an optical waveguide consisting of a highly transparent polymeric material. A thin PMMA cladding with a lower refractive index encircles the PMMA core, causing a total internal reflection, an optical phenomenon, which always occurs when light strikes a medium with a lower refractive index and is reflected to almost full extent. Thus, light cannot leave the waveguide making POF usable for communication technology.

The current surge in POF uses in especially visible in market growth – compared to the market for glass optical fibres, the POF market is booming. While POF technology has been around since the late 70s, using polymer fibres for data communication has been a costly business until the turn of the century when prices for transmitter and receiver modules in the visible wavelength area (400nm to 800nm) declined, making a cost-effective use of POF possible [Da01].

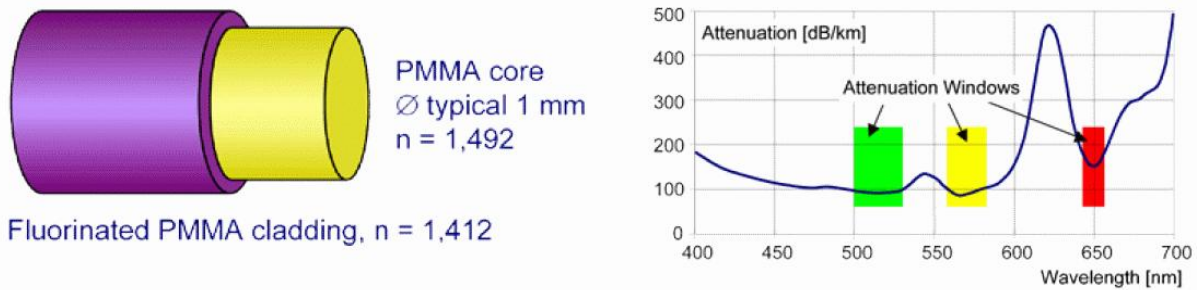

Fig. 1: General structure and of POF (left) and attenuation within the visible spectrum (right) [PO07]

According to a recent study by IGI Consulting [IGI06], the increased interest in POF is mostly due to several current developments in the technical area:

- The demand for cost-efficient high-speed communication technology is increasing
- European automakers have introduced the POF-based MOST-bus[1]
- The 1394b standard has been introduced, increasing the distance between communication nodes to 100m for 3,29 Gbps communication systems
- During the last years, several new POF application fields have been found, including home infotainment, industrial Ethernet, medical technology and sensor technology

Another recent market study, conducted by Harz University itself in 2005 among the members of OPTECNET – the German optical competence networks[2] – shows a clearly risen interest in POF technology. More than 50% of all companies polled are currently in the process of or preparing to expand the use of POF in their own production activities.

Aside from the automotive industry, the industry expected to most heavily shift to POF usage over the next years is the home entertainment sector. A recent market analysis [Ah07] confirms, that although wireless communication systems like WLAN or Powerline Communication have the advantage of relieving the home owner from actually passing any wires, their data rate as well as their technical stability compare to badly against the established Fast Ethernet to make both alternatives viable ones. Lightweight and transparent POF provide home owners an opportunity, to belatedly establish a 100MBit/s data connection without too much effort (because of the extremely simple handling) that is a lot less visible than regular copper cables. Thus, POF has become increasingly interesting for the so-called "last mile" – the last few meters from any city-wide broadband and glass fibre based network to the end user.

The increasing importance of POF and WDM systems makes enabling students and vocational trainees to understand how WDM and/or POF systems are designed, built and maintained a paramount task for universities and other institutions of higher learning or expert vocational training that offer education in optical technology. This includes honing the practical skills of students and vocational trainees and introducing them to concepts such as WDM not only on a theoretical, but also on a practical, "hands-on" level.

---

[1] http://www.mostnet.com
[2] http://www.optecnet.de

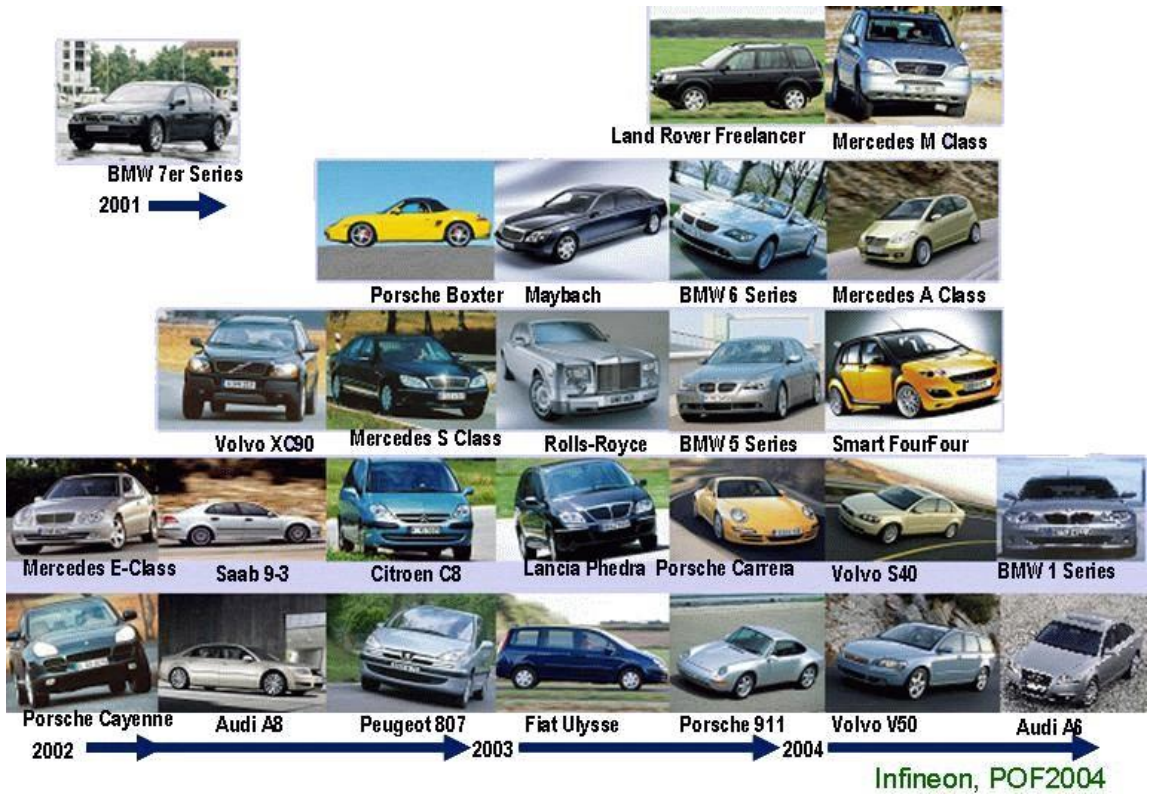

Fig. 2: As of 2004, the MOST bus has already been in use in some of the most recognized modern car types.

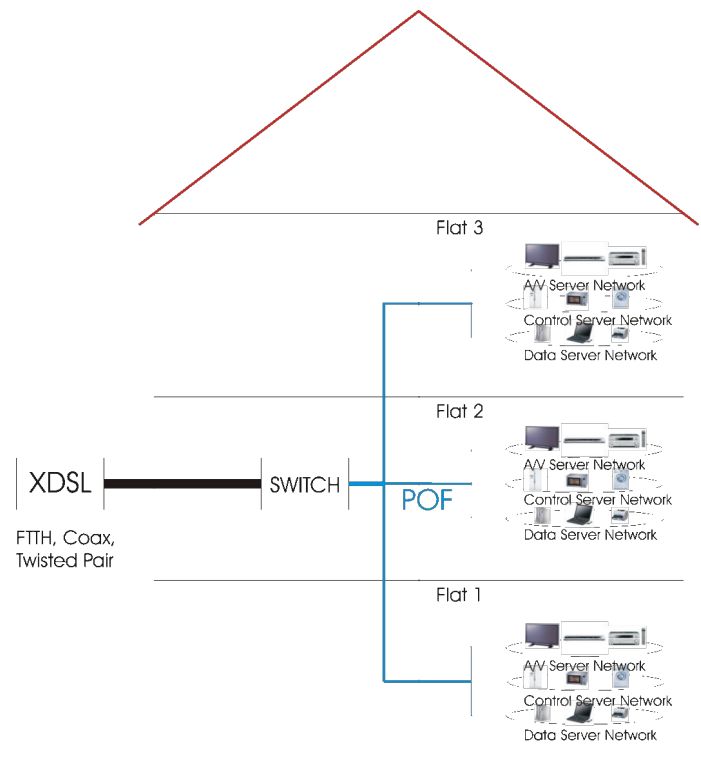

Fig. 3: In-house multimedia infrastructure using POF

In the current academic setting, theory is mostly being taught in the classroom while students gain practical knowledge by performing lab experiments, often using specially designed lab systems. During such experiments, supervising educators have to adapt to the individual learning progresses of single individuals or work groups. In the course of one lab experiment, it is often necessary to quickly and individually rehearse theoretical knowledge or to give problem-specific practical advice. This does not correspond with the idea that students should handle most lab work autonomously as part of the learning experience. By applying the principles of augmented learning to photonics training, contemporary lab work can be brought closer to these ideal conditions.

This paper introduces OPTOTEACH, a newly developed optical teaching system for photonics training in POF data communication and WDM methodology. The paper details the technical layout of the system as well as some of the design concepts behind it. It is then explained, how the system can be augmented with supplemental, interactive software and how this combination of lab system and software creates an effective augmented learning environment.

## 2 Optical Teaching System

OPTOTEACH is the first POF-WDM teaching system, specifically designed to cover a multitude of lab experiments in the field of optical communication technology, e.g. PI curve and bandwidth measurements or analysis of EMF influences. OPTOTEACH systems are exclusively built and distributed by HarzOptics, an optics think tank and research institute associated with the department of Automatation and Computer Science at Harz University in Wernigerode. OPTOTEACH systems are currently being used for educational lab work at Harz University, Braunschweig University, Dresden University, the University of Mannheim and the Federal Centre for Electronics Technology in Oldenburg.

OPTOTEACH systems consist of two video transmitters, one LED and one laser in cw mode and two receivers. The system enables students to transmit two analogous FBAS video signals or corresponding test signals with a maximum bandwidth of 10 MHz. Both transmitters operate within the visible wavelength, which does not only allow OPTOTEACH systems to be built and maintained at reasonable costs, but also provides students with an opportunity to visually experience the WDM effect first hand. The two signals are joined via a conventional Y-coupler developed by Ratioplast Optoelectronics GmbH[3], the separation is effected by a Ratioplast splitter in combination with red and blue colour filters. Signals can be transmitted over various fibre length, covering 5m up to 100m, whereas the fibre itself is interrupted by a micrometer stage, enabling the students to analyse coupling losses with cut or polished fibres as well as lateral and longitudinal misalignments. The general design of the system can be seen in figure 4.

The system gives students an opportunity to perform a multitude of experiments, e.g.:

- PI curve measurements
- Bit Error Rate measurement
- Signal quality tests (eye diagram)
- Measurement of bandwidth and S-parameter
- Analysis of EMF influence on the transmission
- Identification of modulation characteristics (AM, ASK, PCM)
- Attenuation measurements for different fibre lengths (1-100m)
- Attenuation measurements for different wavelengths (490/520/660nm)
- Analysis of the influence of lateral and longitudinal misalignments on the transmission

A more detailed description of the teaching system itself can be found in [Fi06] and [Re06].

---

[3] http://www.ratioplast.com

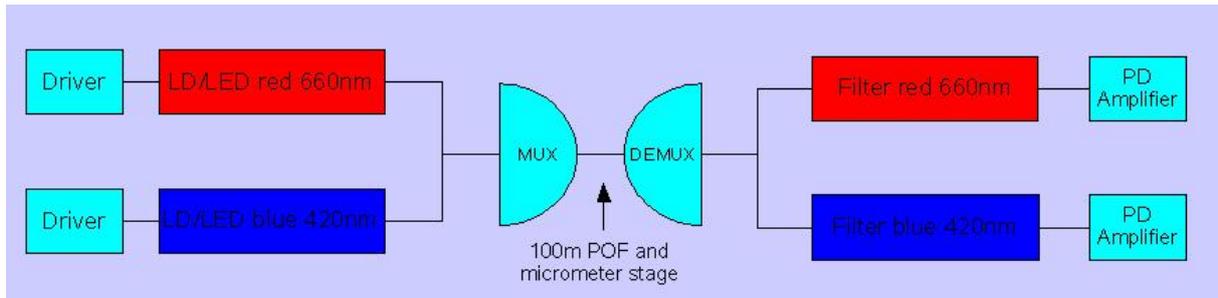

Fig. 4: General technical layout of the OPTOTEACH lab system [Re06]

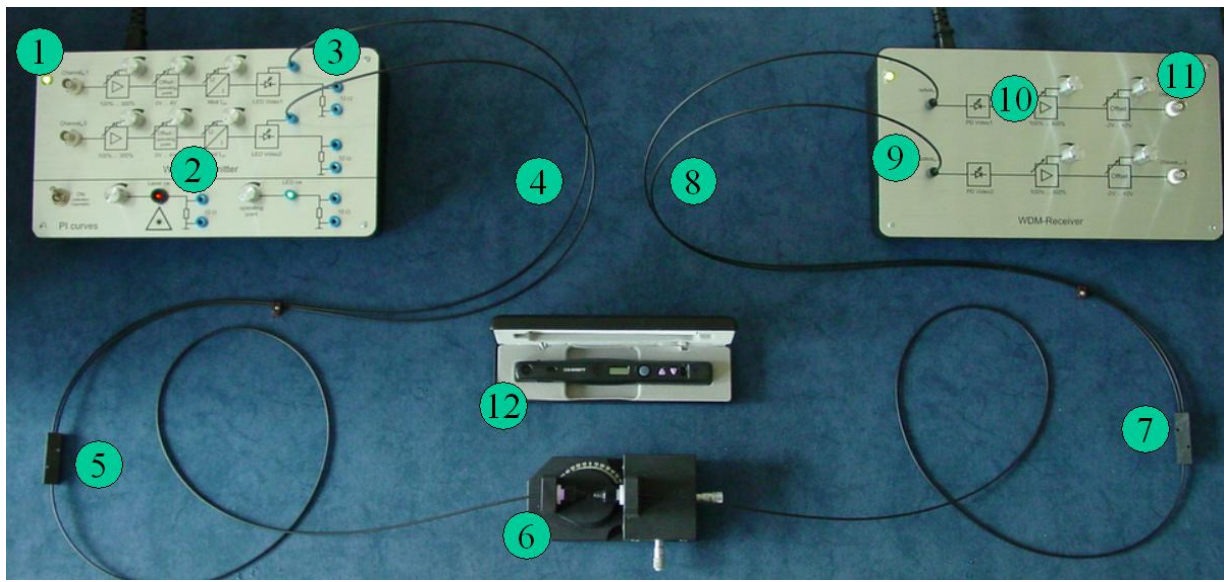

Fig. 5: OPTOTEACH Lab System with (1) BNC-Inputs (2) Potentiometer (3) Optical Outlets (4) Polymer Fiber (POF)  (5) Multiplexer (MUX)  (6) Shifting Table (7) Demultiplexer (DEMUX) (8) Polymer Fiber (POF) (9) Optical Outputs  (10) Potentiometer  (11) BNC-Outputs  (12) Optical Powermeter

### 3 General Software Design

3.1 Basic Requirements

Two basic requirements can be determined for teaching software in general: platform independence and the integration of multimedia content.

The software designer has to make sure, that the software can be used independent from the technical equipment available in the universities or vocational schools. This implies, that platform independent technology such as Java or HTML has to be used at all stages of the software development process.

If HTML is used, if has to be considered, that the terminals used in the educational institutions will differ from each other in browser type and version as well as in screen resolution. Thus, the software has to be thoroughly tested and adapted to the various possible configurations before being released. The development of teaching software for any special combination of operating system, browser type and version as well as screen resolution is economically unsound, because the customers are forced to adapt their technology to the software requirements or be content with a lower quality or hardware-triggered software errors.

Integrating multimedia content into the software application is less of a technical and more of a didactic necessity. The contemporary software user generally expects content to be enhanced with multimedia features and the integration of video films or animations has long been known to be a good practice for activating the user's interest and for making teaching software more appealing [Te00]. Short video sequences and animations that depict can be used to visualize scientific theory as well as depicting actual lab work sequences or experiments. Thus, they can be seen as chapters in a "taped instruction handbook" and an amendment of textual descriptions of experimentation sequences or lab work instructions.

The OPTOTEACH software concept acknowledges these possible problems and depends solely on multimedia technologies, that do not require any plugins (such as animated GIFs) or standard plugins that can reasonably be expected on most of all currently used lab terminals (such as Macromedia Flash).

3.2 Navigation

The direct comparison of online questionnaires in market research and interactive teaching software reveals a common design problem: Should contextual information be placed on one scrollable page or should all content be split into smaller information units that can be displayed on one single screen each [Te00].

If at lot of information is displayed on one single page, the overall theoretical context can be compassed almost instantaneously by the student. This prevents any feelings of being confronted with a seemingly endless number of smaller information screens and allows students to get a quick overview of the entire content and to guess the approximate reading time. Such systems are much less complex – from the programmer's point of view – and are therefore easier and quicker to realize than the programming of a more elaborate system of smaller information screens [Te00]. On the other hand, presenting the entire content of one chapter or the entire proceedings of one experiment can entice students into quickly scrolling through the entire text or completely skip the theory to start with the experiment right away.

The most fundamental benefit of smaller information screens is, that it spares students the discomfort of having to scroll through the information – the navigation is much more concise and brings about a more comfortable software handling. It is also possible to easily integrate interposed control questions between the information screens and to instantaneously validate any given answers, which not only enables the students to get an immediate feedback on their learning efforts but also makes it possible for the software to suggest the targeted repetition of certain theoretical aspects based on the direct evaluation of the answers given.

The information screen option therefore offers a higher level of interactivity as well as enhanced possibilities to evaluate student performance. These advantages and the consequential higher software quality and enriched learning experience outweigh the higher complexity in design and programming. To circumvent the aforementioned feeling of "endlessness", a progress bar can be included, which indicates the remaining number of information screens. Additionally, the average time of completion can be shown at the beginning of each self-contained learning module.

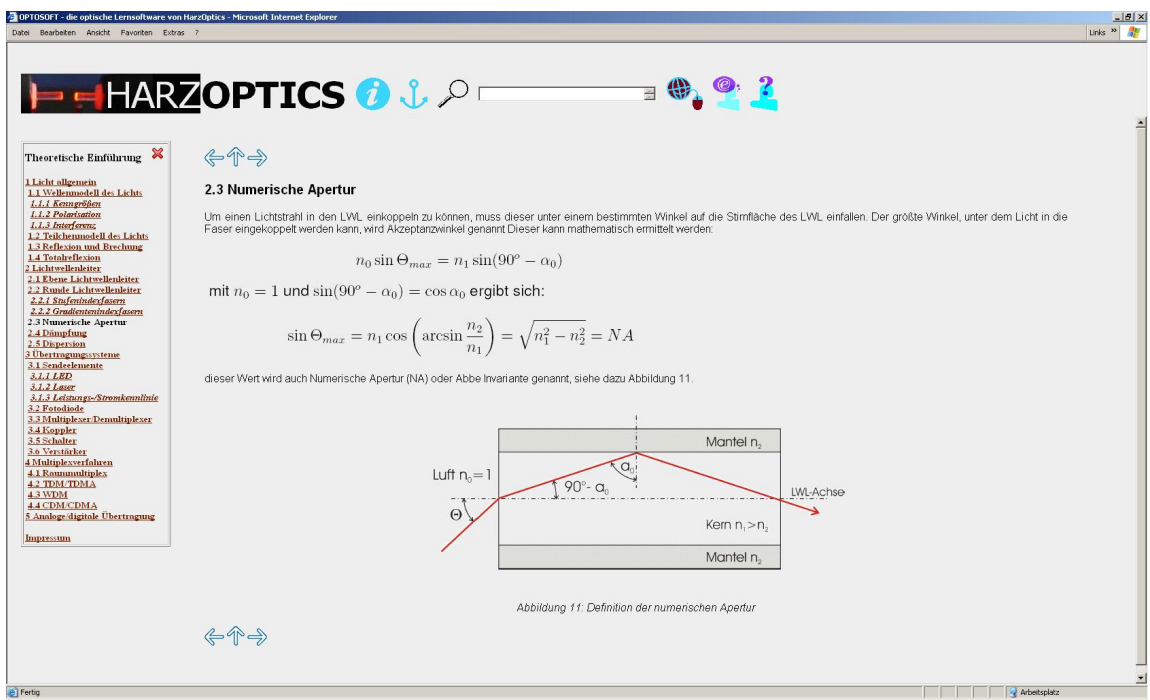

Fig. 6: Screenshot of the current OPTOSOFT beta version (in German language)

The navigation design should allow a comprehensive overview of the theoretical content and enable the student to jump back and forth between the theoretical chapters as well as follow an "ideal learning path". A good example for a practical and concise navigation design is the popular SelfHTML HTML learning software[4]. Another exemplary navigation design can be found on the two online learning sites "Mikro Online"[5] and "Makro Online"[6], developed by Wilhelm Lorenz, professor of macroeconomics at Harz University.

After the navigation design is completed, a pre-test should be arranged along the lines of the conceptual design of any online market research questionnaire. The pre-test allows a testing of the navigation design as well as the general visual impression of the software during a phase of the software development process, in which changes in either the navigational design or the visual presentation are still possible. The basic procedure of such a pre-test can be adopted from market research online pre-testing processes and is described e.g. in [Po98].

## 4 Didactic Concept

4.1 Learning Phases

Because the software aims to support the entire learning process form theory rehearsals to lab-based experiments, it is important to break down the complete process into all methodically different learning phases. Concerning the OPTOTEACH optical teaching system, these phases are already known from the direct practical use of the system in various courses at Harz University:

- Repetition and solidification of theoretical knowledge
- Overview of and support during various lab experiments
- Gathering of measurement data and production of lab protocols

To decide on the ideal didactic concept for the teaching software, the authors extensively researched the various parallels between online collection of market research data, especially via online questionnaires, and lab and/or teaching software within the context of an augmented learning environment. Table 1 contains an excerpt from the list of researched parallels. Similarities were especially appearant concerning the somewhat limited user motivation, which is a problem for market researchers as well as for lab instructors. It is noteworthy that the solution to this problem consists – in both cases – in the introduction of an extrinsic motivational element into the situation, which is known as an incentive in market research terms – and as a grade for students. Both situations demand a certain level of focused mental concentration on the user side, in both cases data is collected and later analyzed and the exact technical configuration of the end unser terminals is unknown to the market research questionnaire designer as well as to the teaching software programmer. In both cases, no specific technological requirements (e.g. operating system, browser type, browser version or number of additional plugins needed) can be made without excluding potential users. Many more parallels can be found, e.g. concerning the average time needed to complete a typical online questionnaire or an average learning module.

| Feature | Online Questionnaire | Lab / Teaching Software |
| --- | --- | --- |
| Level of Motivation | Low or very low | Partially low |
| Source of Motivation | External (Incentives) | External (Grades) |
| Focus of Participants | Usually high | Mostly high |
| Data Analysis Method | Analysis of given answers | Evaluation / Grading |
| Programming | Java, HTML, CGI | Java, HTML, CGI |
| User-side IT Technology | Manifold technology | Manifold technology |
| Average Duration | 20-30 Minutes | 30-40 Minutes |

Tab. 1: Parallels between the online collection of market research data through questionnaires and the use of teaching software within or outside an augmented learning environment (excerpt)

---

[4] http://www.selfhtml.org
[5] http://www.mikroo.de
[6] http://www.makroo.de

Because of these parallels, it seems prudent to utilize already existing scientific research on the creation of ideal conceptual designs for market research questionnaires, especially the existing Best Practice frameworks, in the development of teaching software. A thorough review of contemporary online market research methodology also confirms other findings about the ideal design of teaching software: direct feedback and a high degree of interactivity can trigger a heightened involvement on the user side, the (careful and spare) use of high quality multimedia elements helps to keep up the user's attention and the best possible solution to present a larger number of questions (or other content) is splitting them up in screen-sized information modules.

The application of the most important guidelines in contemporary online market reasearch questionnaire design (especially the research of [Te00], [Dr03], [Bö99] and the most recent [We05]) lead to the four basic software modules pictured in fig. 7.

These four modules are: The continuous repetition and cementation of theoretical knowledge about various aspects of optical technology, the user-controlled exploration and self-testing of this theoretical content utilizing interactive graphs and modules such as multiple-choice questionnaires, the customisable help and support of lab experiments and the option of generating and saving PDF[7] protocols with measurement data and student answers to theoretical questions as a data base which can be utilized by the lecturer for grading purposes.

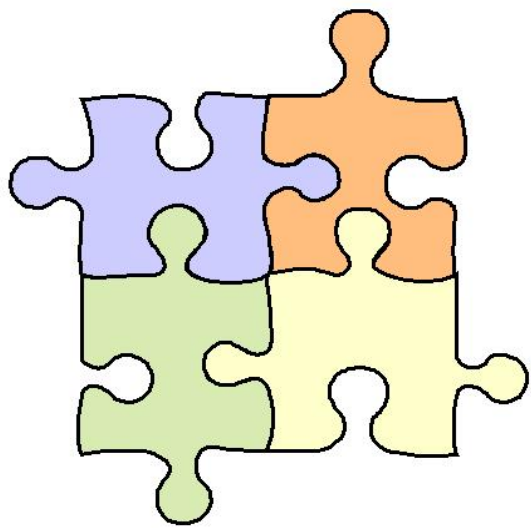

Fig. 7: Overview of the four OPTOTEACH basic software elements

The repetition and consolidation of theoretical knowledge is not confined to one theoretical module which students have to complete before a lab experiment can begin, instead, theoretical knowledge is repeated throughout the entire experimentation process and the interpretation of resulting measurement values. The hypertext-character of HTML allows the implementation of this idea into the software, because students can use embedded hyperlinks on important science terms to jump to corresponding theoretical context and then back to the current experiment. A permanently accessible glossary with an integrated search function makes it even easier for students to gain access to important theoretical knowledge. The synergy-effect that results from the interconnection of hands-on experimental lab work and understandable overviews of theoretical basics can thus be utilized most efficiently.

Because the overall design (the single information screen model mentioned above) allows the easy implementation of interposed control questions, this additional control method, which is typically not part of typical educational lab work programmes, can be integrated into the experimental workflow. It is up to the lecturer or lab administrator to decide, whether these knowledge checks are mandatory or optional for students.

The chaperonage of the individual student during lab experiments – the only part of the teaching software that is currently still under development – forms the core of the augmented learning environment. Parallel to the actual experimental performance students will be able to inform themselves about the general layout of the experiment, get work instructions from the computer and enter their measurement data, whereas average measurement data and other intermediate results provide the opportunity to continuously check whether the experiment is conducted correctly. Thus, students will be enabled to detect any discrepancies in their measurement data at an early stage and therefore check their own results, which is almost impossible to realize in a traditional lab environment. In the event of perceptible deviations of the experimental results from the ideal results, a set of multiple-choice questions will allow the student to identify probable causes of the discrepancy and gather instructions for correcting any possible mistakes. The lecturer only has to get involved into a particular experiment if this help system does not provide the solutions needed to achieve the expected results.

---

[7] PDF (Portable document File) is a registered trademark of Adobe Inc.

The heightened level of autonomy alleviates student-controlled lab work and supports the pedagogical concept behind of enabling students to advance their practical skills as well as their theoretical knowledge more or less autonomously in a self-controlled environment.

The acquisition of measurement data and the compilation of lab protocols will also be implemented, whereas the software covers all four learning phases. Via a HTML form field, students can enter measurement data as well as textual answers to theoretical questions and questions about completed experiments. This does not only represent a significant assistance for students but also for lecturers and lab administrators who will be disburdened from deciphering bad handwritings and searching for lost sheets of paper.

4.2 Dimensions of Teaching Software

Teaching software, like the OPTOSOFT software presented in this paper, is basically defined through the three dimensions of interactivity, adaptivity and controllability [DE01].

According to [Ke98], interactivity can be seen as a mostly technical dimension: When working in an interactive medium, the user – in this case the optics student – has unrestricted and self-controlled access to multimedia information. The interactivity allows the active processing of teaching content by the student, who has the ability to influence the selection and the sequence of content at least partially [Ja00]. Within OPTOTEACH, interactivity is provided via the easy-to-use navigation, which allows almost unrestricted access to all content modules. Students can forgo the recommended "learning path" and navigate freely through the software.

Adaptivity is defined as the extent to which users are allowed to customize any given software [DE01]. OPTOSOFT allows students to adjust the software to the preferred working speed, repeat complex passages at will or self-check the comprehension via multiple choice questions. When the software is used as a lab companion, the speed of instructions and recommendations is adjustable to the actual experimental progress, likewise in the acquisition of measurement data and the compiling of lab protocols, meaning the dimension of adaptivity is distinctive throughout all four phases described in 4.1.

According to [DE01], controllability does not refer to the control of the lecturer or lab administrator over the student but to the control of the student over the learning process. In computerized learning environments, the controllability increase with the extent to which non-linear navigation is implemented, meaning the less restricted the user is, the higher is the controllability of the software [Ne00]. Because the technical basis of OPTOSOFT is HTML, the hypertext-functionality allows a nearly completely unrestricted navigation throughout most of the learning modules. The only restrictions will be implemented into the lab protocol compilation process, because the electronic documents generated in this process may provide a basis for student grading.

Because of that, students will only be able to access the protocol editor after an experiment has been completed instead of being able to directly jump into the protocol compilation process. Furthermore, the PDF files generated by OPTOSOFT will exhibit a time stamp unchangeable by student changes in the protocol editor.

## 5 First Experiences

At present, more than half of the teaching software is completed, with the theoretical modules being already fully functional. The software will soon be undergoing vigorous beta testing at the Harz University photonics labs. The first student and lecturer feedbacks have been unanimously positive, welcoming the introduction of the multimedia element and the easy-to-navigate glossary to the lab. Several technical problems have been asserted and will be rectified before the release of the pre-beta-version.

One of the more interesting results of the first feedback evaluation is the clear demand for more multimedia elements to be included in the final version, directly connected to the general wish for a more colourful and less conservative visual design. While these wishes can certainly be implemented in the pre-beta-version, it is important not to overload the software with colours and multimedia elements, preserving the scientific image.

## 6 Conclusion and Outlook

At this stage, the fully functional OPTOSOFT version 1.0 is expected to be complete in late 2007, so that universities and vocational schools can start using the software in class and lab work not later than early 2008. The current version already includes the complete theoretical learning modules, the knowledge base, the link list, most of the multiple choice tests and interactive modules as well as some of the lab companion modules. Other lab companion modules, lab movies and a fully functional version of the protocol generator are still in development. The OPTOTEACH lab teaching system also described in this paper, has already been successfully introduced to the German education market and, as of early 2007, is being used in more than half a dozen universities nationwide.

Because the adoption of the augmented learning idea to photonics lab work is new and untested, the authors are very curious about the feedback of the first classes and lecturers that will start working with the software in early 2008. A quality feedback system, that will allow students and lecturers alike to communicate their experiences and critique points, is already in an early set-up phase. Aside from the mostly subjective impressions of students and lecturers, this feedback system will also gather more objective data such as average grading before and after the introduction of the software as well as the results of the teaching quality evaluation during the introductory period. A quality control and a version management system will ensure, that didactic and technical change requests are collected, evaluated, implemented or not implemented and archived for further evaluation.

The continuous further development of teaching system and teaching software as well as the complete documentation of this development process will result in a comprehensive catalogue of specifications and problems in the practical use of an interactive and multimedia lab companion software in the context of a photonics augmented learning environment. The authors expect to publish a revised edition of this catalogue as a basis for discussion about the optimal way of introducing augmented learning to photonics training.

All public and private institutions of vocational and higher education are invited to participate in this project.

## 7 Acknowledgement

The authors want to tank the German Federal Ministry for Education and Research as well as the state government of Saxony-Anhalt, who funded and supported the work described in this paper.

## Sources